\documentclass[jcp,aps,final,preprint,groupedaddress,showpacs,floatfix,aip,reprint]{revtex4}

\input pstricks \input epsf
\usepackage{graphicx}
\begin{document}

\def\beq{\begin{equation}}
\def\eeq{\end{equation}}

%\preprint{}

\title{Molecular dynamics simulations of D$_{2}$O ice photodesorption}

\author{C. Arasa}
\affiliation{Leiden Observatory, Leiden University, P. O. Box 9513, 2300 RA Leiden, The Netherlands and \\ Gorlaeus Laboratories, Leiden Institute of Chemistry, Leiden University, P. O. Box 9502, 2300 RA Leiden, The Netherlands}

\author{S. Andersson}
\affiliation{SINTEF Materials and Chemistry, P.O. Box 4760, 7465 Trondheim, Norway and \\ Department of Chemistry, Physical Chemistry, University of Gothenburg, 41296
Gothenburg, Sweden}

\author{H. M. Cuppen}
\affiliation{Leiden Observatory, Leiden University, P. O. Box 9513, 2300 RA Leiden, The Netherlands}

\author{E. F. van Dishoeck}
\affiliation{Leiden Observatory, Leiden University, P. O.  Box 9513, 2300 RA Leiden, The Netherlands}

\author{G. J. Kroes}
\affiliation{Gorlaeus Laboratories, Leiden Institute of Chemistry, Leiden University, P. O. Box 9502, 2300 RA Leiden, The Netherlands}

\date{\today}% It is always \today, today,
             

\begin{abstract}           
Molecular dynamics (MD) calculations have been performed to study the ultraviolet (UV) photodissociation of D$_{2}$O in an  amorphous  D$_{2}$O ice surface at 10, 20, 60, and 90~K,  in order to investigate the influence of isotope effects  on the photodesorption processes.  
As for H$_{2}$O, the main processes after UV photodissociation  are trapping and desorption of either fragments or D$_{2}$O molecules. 
Trapping mainly takes place in the deeper monolayers of the ice, whereas desorption occurs in the uppermost layers. There are three desorption processes: D atom, OD radical,  and D$_{2}$O molecule photodesorption. D$_{2}$O  desorption takes places either by direct desorption of a recombined D$_{2}$O molecule, or when  an energetic D atom produced by photodissociation kicks a surrounding D$_{2}$O molecule out of the surface by transfering part of its momentum. 
 Desorption probabilities are calculated for photoexcitation of D$_{2}$O in the top four monolayers and compared quantitatively with those for  H$_{2}$O obtained from   previous MD simulations of UV photodissociation of amorphous water ice   at different ice temperatures [Arasa $et~al.$, J. Chem. Phys. {\bf{132}}, 184510 (2010)]. 
 The main conclusions are the same, but the average D atom photodesorption probability is smaller than that of the H  atom (by about  a factor of 0.9) because D has lower kinetic energy than H, whereas the average OD radical photodesorption probability is larger than that of OH (by about a factor of 2.5--2.9 depending on ice temperature) because OD has higher translational energy  than OH  for every ice temperature studied.
 The average D$_{2}$O photodesorption probability is larger than that of H$_{2}$O (by about a  factor of 1.4--2.3 depending on ice temperature),  and this is entirely   due to  a larger contribution of  the D$_{2}$O kick-out mechanism. 
This is an isotope effect: the kick-out mechanism is more efficient for D$_{2}$O ice, because the D atom formed after D$_{2}$O photodissociation has a  larger momentum  than  photogenerated  H atoms from H$_{2}$O, and D transfers momentum more easily to D$_{2}$O than H to H$_{2}$O. 
The total  (OD + D$_{2}$O) yield has been compared with experiments and the total (OH + H$_{2}$O) yield from previous  simulations.
We find better agreement when we compare experimental yields with calculated yields for D$_{2}$O ice than when we compare with calculated yields for H$_{2}$O ice.

\end{abstract}
  
  \maketitle %\maketitle must follow title, authors, abstract and \pacs

\section{Introduction}

The formation of  molecules in the interstellar medium (ISM) can proceed through several kinds of reactions. Surface reactions on  nano- to micrometer sized particles are thought to play a key role in the formation of molecules in the ISM \cite{Herbst2009}. Dust grains in the ISM consist of a core 
of silicates  and carbonaceous components and, in dense clouds, these dust particles can be covered by icy mantles. The icy mantles contain mainly 
H$_{2}$O, but also traces of other molecules  (e.g., CO, CO$_{2}$, NH$_{3}, $CH$_{4}$, among others) \cite{Tielens1991, Boogert2008}. 
Observed infrared (IR) spectra  reveal that H$_{2}$O and CO are the most abundant molecules in the icy mantles in the ISM \cite{Tielens1991, Tanaka1994, Chiar1995, Gibb2000, Gibb2004, Pontoppidan2003, Pontoppidan2006,  Boogert2008, Zasowski2009}. 
Recent ground and space based observations have also  detected heavy water (D$_{2}$O) in the ISM \cite{Butner2007, Vastel2010}.

Ultraviolet (UV) irradiation of an icy grain can photodissociate water molecules and   cause desorption of the ice. The flux of UV photons in the ISM is  low \cite{Herbst2009, Ehrenfreund2003, Garrod2006} compared with the lamp UV photon flux used in the laboratories (which varies between, e.g., (1.1--5.5)$\times$10$^{13}$ photons cm$^{-2}$s$^{-1}$ \cite{Oberg1}),  with photon fluxes  of the order of 10$^{3}$ photons cm$^{-2}$s$^{-1}$, which is equivalent to roughly one incident photon per month per grain.
The  photodissociation dynamics is typically computed over a picosecond time scale, and hence, the photodissociation by one incident photon is completed before the next photon arrives at the ice. The energies of the incident photons $\sim$6--13~eV \cite{Kobayashi1983, vanDishoeck1988, Boogert2004} cover the first  absorption bands of water ice. 

Photodissociation and photodesorption of water in ice  are of interest to understand astronomical observations of gas-phase water in cold clouds \cite{Jack1988, Knacke1991, Cernicharo1990, Gensheimer1996, vanderTak2006, Kaufman2008, Hollenbach2004, Hollenbach2009}, and also because  the photoproducts  (H and OH) can proceed to react with co-adsorbed species,  which may lead to the formation of more complex molecules \cite{dHendecourt1982, Garrod2006}.
In addition, the process is interesting from a fundamental chemical physics point of view. Most studies of photodissociation processes of molecules on surfaces have focused on (sub)monolayers  of species on mostly metallic surfaces, not on the thick ($\sim$100 monolayer (ML)) ices found in interstellar space. Very different processes can occur in this case.

Several experiments on UV irradiation of amorphous and crystalline    H$_{2}$O ice  \cite{Westley1995a, Westley1995b,Ghormley1971, Gerakines1996, Yabushita1, Oberg1, Yabushita2006, Yabushita2008, HamaAug2009, HamaSet12009, HamaSet22009, Hama2010} and  D$_{2}$O ice \cite{Watanabe2000,Oberg1,Hama2010} have been carried out using different analysis techniques and different light sources.

In order to obtain insight into the basic molecular processes, the photodissociation of H$_{2}$O molecules in amorphous and crystalline  ice in the temperature range 10--90~K has  been studied  using molecular dynamics (MD) simulations \cite{Andersson2005, Andersson2006, Andersson2008, Arasa2010}.  The most important photodesorption mechanism after photodissociation of water   in the top three MLs  of the ice surface is H atom photodesorption, followed by OH radical photodesorption, and   H$_{2}$O molecule photodesorption. 
The calculated H$_{2}$O photodesorption probability
is due to two mechanisms.  (1) The direct mechanism: H and OH recombine after H$_{2}$O photodissociation  to form H$_{2}$O, which eventually desorbs. (2) The kick-out mechanism  \cite{Andersson2006,Andersson2008}:  an energetic H atom released after photodissociation kicks out one of the surrounding water molecules by a  transfer of momentum.

Many key experiments studying water ice photodesorption have been performed for D$_{2}$O rather than H$_{2}$O.
In contrast, the MD simulations  have so far been carried out only for H$_{2}$O. In order to better compare with experiments and to identify isotope effects on the
photodesorption processes, we present here results of MD simulations
of the UV photodissociation of amorphous D$_{2}$O ice at different temperatures.

In Sec.~\ref{sec:methods} we present the methods used in this study, in Sec.~\ref{sec:results} the main results in comparison with previous UV photodissociation of amorphous H$_{2}$O ice results are presented, and in Sec.~\ref{sec:conclusions} the concluding remarks are given.

\section{Methods}
\label{sec:methods}

\subsection{Potentials}
\label{ssec:Potentials}

The total analytical potential energy surface (PES) for the ice in the photodissociation calculations is the same as in our previous studies and can be written as follows:

\begin{equation}\label{eq1}
V_{\rm{tot}} =V_{\rm{ice}}+V_{\rm{H_{2}O^{*}-ice}}+V_{\rm{H_{2}O^{*}}} 
\end{equation}\\

The first term describes the intermolecular interactions between the H$_{2}$O molecules inside the ice excluding the  H$_{2}$O molecule that is photoexcited. These interactions are described by the TIP4P potential \cite{TIP4P} with all molecules kept rigid.

The second term refers to the intermolecular interactions of the photoexcited molecule, which is treated as fully flexible, with the rigid ice molecules and the third term is the intramolecular potential of the photoexcited molecule. These
potential terms also cover all interactions involved in the dissociation and
possible recombination of the excited molecule. The potentials are exactly the same
as previously used for photodissociation of H$_{2}$O ice. All details of the potentials and the functions used to switch between different potentials are given in Ref.~\cite{Andersson2006, Arasa2010}.

\subsection{Amorphous ice surface}
\label{ssec:Amorphous}

To study the UV photodissociation of D$_{2}$O ice, we simply changed the mass of the H atom to that of the D atom.  (Of course, all the interactions that take place in and on  the ice during the photodissociation  are described with the same potentials employed for the UV photodissociation of  H$_{2}$O ice (Eq.~\ref{eq1})).

Crystalline and amorphous D$_{2}$O ice surfaces were constructed  using the MD method \cite{Allen1987}, and using the same procedure and  cell parameters  employed before to model H$_{2}$O ice  \cite{Andersson2005, Andersson2006, Andersson2008, Arasa2010}.

Starting from the normal hexagonal ice (I$_{\rm{h}}$) crystalline ice configuration (containing 8 bilayers (BLs) (16~MLs) with 60 (30) molecules in each ML), the amorphous ice surface was set up at 10, 20, 60, or 90~K  using the `fast quenching' method \cite{Essmann1995, AlHalabi2004a, AlHalabi2004b}. Further details can be found in our previous studies \cite{Andersson2006, Andersson2008, Arasa2010}.
Since the resulting amorphous ice surface has a more irregular bonding structure than the crystalline ice surface  \cite{AlHalabi2004a, Andersson2006} assigning molecules to MLs is not straightforward \cite{Arasa2010}.
 In our most recent sudy \cite{Arasa2010}  a new definition of ML (binning method 2) was tested and shown  to be a more realistic way to assign molecules to MLs.
  This binning (method 2) is used in this study and it consists of choosing a molecule and finding the first 23 closest molecules in terms of ($x$, $y$) coordinates. This leads to 24 molecules       that are divided in 12 bins of two molecules each,  based on their   $z$ center of mass coordinates. The 12 bins represent the top 12~MLs of the ice in which the molecules are allowed to move.

\subsection{Initial conditions and dynamics} 
\label{ssec:Initial}

For each of the top four MLs, all the molecules were chosen to be photodissociated and for each molecule 200 different initial configurations were considered.   To initialize the trajectories \cite{Wigner2}, a Wigner phase-space distribution function \cite{Wigner1} fitted to the ground-state vibrational wavefunction  of gas-phase water is used. In the case of heavy water  the trajectories are  initialized by using the corresponding Wigner distribution of gas-phase heavy water, which has the same functional form  as for  gas-phase water (Eq.~5.13 in Ref.~\cite{Wigner1}), but with  $\alpha_{\rm{D_{2}O}}$ = $\sqrt{2}$ $\alpha_{\rm{H_{2}O}}$ (Eq.5.16 in Ref.~\cite{Wigner1}).
The initial coordinates and momenta of the atoms from the dissociating molecule are sampled using a Monte Carlo procedure. Then, a Franck-Condon excitation is performed and the system is put on the first electronically excited state, on the DK $\rm{\tilde{A}^{1}B_{1}}$ PES  \cite{DK1, DK2,DK3}.

The excitation energies are calculated by computing the energy difference between  a D$_{2}$O ice with an excited molecule and one with a ground state D$_{2}$O molecule (both molecules with the same coordinates). The calculated D$_{2}$O amorphous ice spectrum is  shifted 0.02~eV with respect to that for  H$_{2}$O amorphous ice \cite{Andersson2006}.

To simulate the dynamics of a photodissociation event, Newtons's equations of motion are integrated in time  with a time step of 0.02~fs and a maximum time of 20~ps.    
The stop criterion and the six final outcomes after UV photodissociation of D$_{2}$O are analogous to those described previously for H$_{2}$O photodissociation \cite{Andersson2005, Andersson2006, Andersson2008, Arasa2010}:
(1) desorption of  D while OD is trapped inside or on the ice, (2) desorption of OD  while D is trapped inside or on the ice, (3) desorption of both D and OD, (4) D and OD  are both trapped inside or on the ice, (5) D and OD recombine and form a D$_{2}$O molecule which either desorbs or (6) is trapped inside or on the ice.
Besides these six outcomes,  
an additional channel is possible  where  D$_{2}$O desorbs through the so-called `kick-out' mechanism \cite{Andersson2006, Andersson2008}.  This occurs when a molecule  desorbs from the ice by momentum transfer  from an energetic D atom resulting from photodissociation of a neighbouring photoexcited molecule.

We calculate the probabilities $P^{\rm{i}}$ of the outcomes  per absorbed UV photon in a specific ML $i$ and its standard errors ($\epsilon^{\rm{i}}$=$\sqrt{P^{\rm{i}}\cdot(1-P^{\rm{i}})/N}$, where $N$ is the  total number of trajectories simulated  in  ML~$i$) at all ice temperatures for the top four MLs (the error bars in the figures and tables correspond to  66$\%$ confidence intervals).
However, not all of the UV photons that arrive at the ice are absorbed in these monolayers.  Andersson $et.~al.$ \cite{Andersson2008} 
estimated the absorption probability per ML ($P_{\rm{abs}}^{\rm{ML}}$)
to be about 7$\times$10$^{-3}$  using an absorption cross section of  about  6$\times$10$^{-18}$cm$^{-2}$  (for more details we refer to the Appendix~A in Ref.~\cite{Andersson2008}).
In the case of heavy water amorphous ice we have assumed the same $P_{\rm{abs}}^{\rm{ML}}$.
The total photodesorption yield ($Y$) can be calculated 
from the calculated photodesorption probabilities per absorbed UV photon in a specific ML $i$, by multiplying this probability  $P_{\rm{des}}^{\rm{i}}$  with the probability that the photon makes it to ML~$i$ and the probability that the photon is absorbed in a given ML   ($P_{\rm{abs}}^{\rm{ML}}$), and summing the resulting yields per ML over the considered MLs. This is summarized in the following equation \cite{Arasa2010}:

\begin{equation}\label{eq2}
Y=\sum_{i=1}^{n}P_{\rm{des}}^{i}\cdot(1-P_{\rm{abs}}^{\rm{ML}})^{i-1}\cdot P_{\rm{abs}}^{\rm{ML}}
\end{equation}\\

\section{Results and discussion}
\label{sec:results}

\subsection{D atom  photodesorption}
\label{ssec:Datom}

The probabilities of all the different outcomes following photoexcitation of one molecule in the ice have been calculated, but we only report those concerning photodesorption (i.e., outcomes 1, 2, 3, and 5) and for the top 4~MLs because,  according to previous MD simulations at different ice temperatures \cite{Andersson2006, Andersson2008},  the photodesorption mainly takes place after photoexcitation in these monolayers.
Thus, the outcome  probabilities strongly depend on the monolayer in which the photoexcited molecule is initially located:  the photoexcitation in the top MLs   leads mainly to photodesorption, while deeper into the ice it leads to trapping \cite{Andersson2005, Andersson2006, Andersson2008, Arasa2010}.

\begin{figure}[htbp]
\begin{center}
\includegraphics[width=8cm]{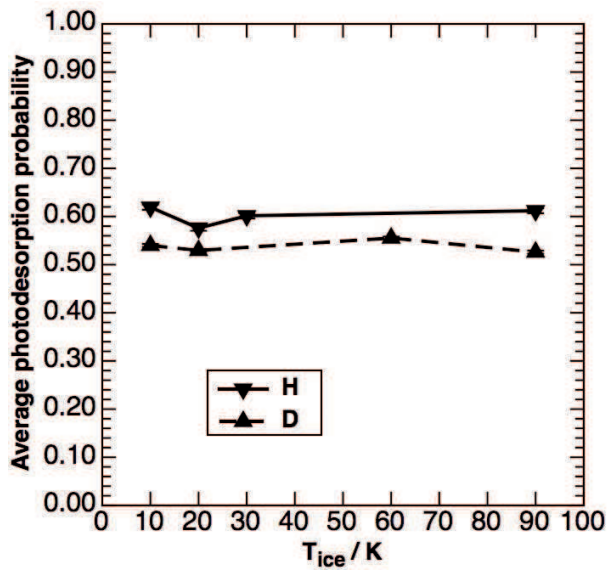}
\end{center}
\caption {{ The probability of D atom (dashed line) and H atom (solid line) photodesorption averaged over the top four MLs per absorbed UV photon is shown as a function of ice temperature. H atom results from Ref.~\cite{Arasa2010}.}}
\label{Figure1}
\end{figure}

The average D atom photodesorption probability and the average H atom photodesorption probability \cite{Arasa2010} taken over the top four monolayers (e.g., $\textless$$P_{\rm{Hdes}}$$\textgreater$=$\sum_{\rm{i}=1}^{4}P_{\rm{Hdes}}^{\rm{i}}$/4) are plotted in Fig.~\ref{Figure1}.   The average H photodesorption probability is somewhat larger than that of D, by about 6~$\%$ for all $T_{\rm{ice}}$ studied. 
The total  deuterium (hydrogen) atom photodesorption probability is calculated by summing over two different processes: one in which the D (H) atom desorbs while the OD (OH) stays trapped in the surface, and one in which both photofragments desorb from the ice surface (outcomes 1 and 3).

\begin{figure}[htbp]
\begin{center}
\includegraphics[width=10cm]{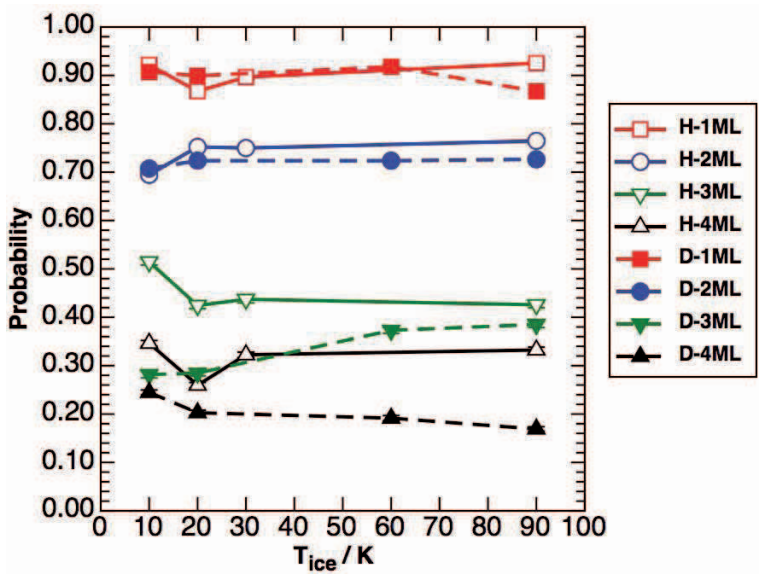}
\end{center}
\caption {{ Total probability of D atom (dashed line) and H atom (solid line) \cite{Arasa2010} photodesorption (per absorbed UV photon) versus temperature,  for the uppermost four MLs.}}
\label{Figure2}
\end{figure}

The deuterium  and hydrogen atom photodesorption probabilities   are higher than the OD (OH) and D$_{2}$O (H$_{2}$O) photodesorption probabilities in the uppermost monolayers of the D$_{2}$O (H$_{2}$O) amorphous ice  (see Fig.~\ref{Figure2} where the H and D photodesorption probabilities are displayed versus ice temperature and ML).
This is because  D and H atoms are smaller and are formed with higher kinetic energies immediately after D$_{2}$O and H$_{2}$O photodissociation, which  
facilitates the desorption of these atoms  \cite{Andersson2005, Andersson2006, Andersson2008, Arasa2010}.

The average photodesorption probability of D  is smaller than that of H mostly because  
the   probabilities of D atom  photodesorption in the third and fourth monolayers (Fig.~\ref{Figure2})  are smaller than those for the H atom.
This trend is expected because D is heavier than H. Therefore, the efficiency of energy transfer between D and D$_{2}$O molecules is larger than the corresponding efficiency between H and  H$_{2}$O molecules.
If the photoexcited molecule is isolated (i.e., in the absence of the surrounding ice), the initial kinetic energy in which D and H atoms are formed after D$_{2}$O and H$_{2}$O photodissociation should be similar in order to achieve energy conservation. 
But,  in the presence of ice, D atoms lose more kinetic energy than  H atoms when they  interact with the surrounding molecules due to the larger  efficiency of energy transfer, and they are therefore less able to penetrate the ice when moving through the upper layers.

The dependence on ice temperature is negligible. The average D atom photodesorption probability (Fig.~\ref{Figure1}) is almost constant at $\sim$54~$\%$.  However, this  probability depends on the ML where the photoexcited  molecule was initially located (Fig.~\ref{Figure2}).  In the top two MLs the probability is high ($\sim$90~$\%$ to $\sim$70~$\%$), but it drops in the third ML and further below, because  other processes such as  trapping are in competition.
Trapping becomes important  because deeper in the ice the structure is more closed  and  the molecules from the ice above can impede the D atom from reaching the ice surface.

D atoms travel through the D$_{2}$O ice at 90~K by an  average distance of  8.4~\AA~before they become trapped, whereas  H atoms travel around 9.1~\AA. The OD and OH  radicals travel  2.2~\AA~and  1.9~\AA, respectively. The recombined D$_{2}$O and H$_{2}$O move on average a distance of  1.8~\AA~and 2.0~\AA, respectively.
Therefore, the mobility of the photofragments inside the ice  is slightly affected by the mass of the photofragments: H atoms move further than D atoms until they become trapped, because H  is lighter than D. The maximum  distances travelled, which  are about tens of angstroms should enable  reaction with other species trapped in the ice. This could explain  the formation of more complex molecules in the ISM.

\subsection{OD radical  photodesorption}
\label{ssec:OD}
\begin{figure}[htbp]
\begin{center}
\includegraphics[width=8cm]{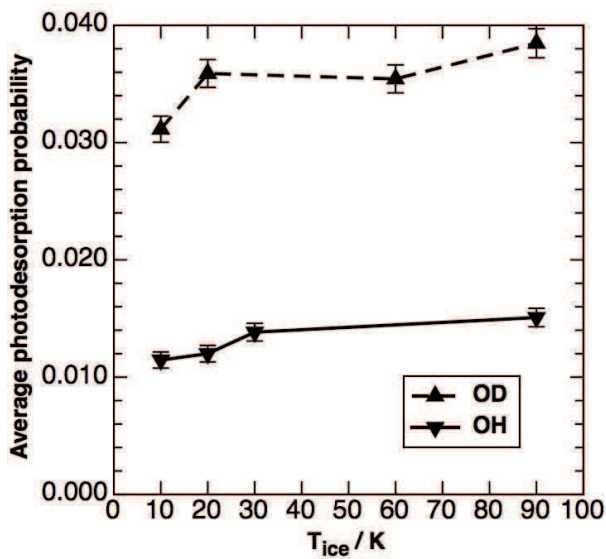}
\end{center}
\caption {{ The probability of OD radical (dashed line) and OH radical (solid line) photodesorption averaged over the top four MLs per absorbed UV photon is shown as a function of ice temperature. OH radical results from Ref.~\cite{Arasa2010}. }}
\label{Figure3}
\end{figure}

 The second main photodesorption mechanism  in the uppermost MLs of the ice is OD photodesorption. 
Fig.~\ref{Figure3} shows that the average of the OD photodesorption probabilities taken over the top 4~MLs is larger than that for  OH for all ice temperatures studied. To calculate the OD (OH) photodesorption probabilities we have summed over the probabilities of two pathways: the probability of the channel in which the OD (OH) radical desorbs while the D (H) atom remains trapped in the ice surface, and the probability of the channel in which both photofragments leave the ice surface (outcome 2 and 3, respectively).

\begin{figure}[htbp]
\begin{center}
\includegraphics[width=10cm]{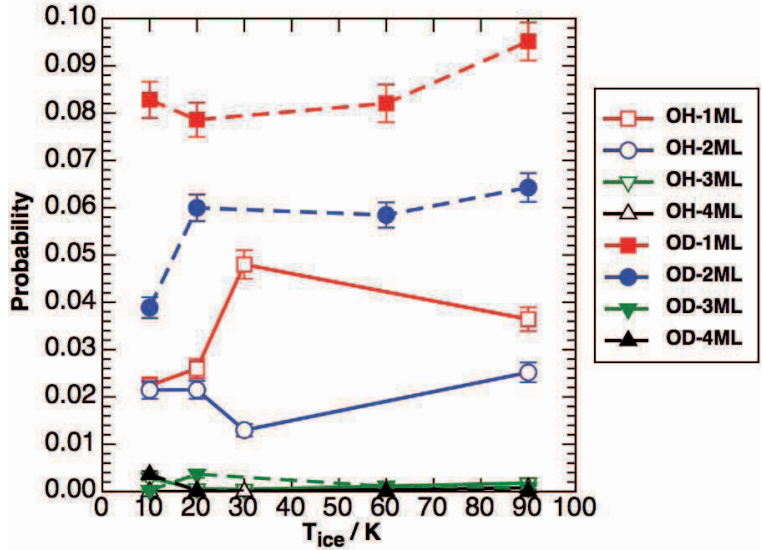}
\end{center}
\caption {{ Total probability of OD radical (dashed line) and OH radical (solid line) \cite{Arasa2010} photodesorption (per absorbed UV photon) versus ice temperature,  for the uppermost four MLs.}}
\label{Figure4}
\end{figure}

The   probabilities  per monolayer are plotted in Fig.~\ref{Figure4} versus ice temperature.
Those for OD  are much larger than those for  OH  \cite{Arasa2010} in the top two monolayers, which gives rise to a larger total average OD photodesorption probability (Fig.~\ref{Figure3}). 
In the absence of the surrounding ice,  the water fragments (D and OD, H and OH) have to obey momentum conservation ($p_{\rm{X}}$=$-p_{\rm{OX}}$, $X$=H, D) and energy conservation ($E_{\rm{X}}$+$E_{\rm{OX}}$=$E_{\rm{X_{2}O}}$=$E$,    $E$ being the initial available energy $E_{\rm{exc}}$$-$$E_{\rm{diss}}\rm{(X_{2}O)}$, $X$=H, D;  the excitation energy  $E_{\rm{exc}}$ is in  the range  7.5--9.5~eV with a peak at 8.6~eV, and the dissociation energy $E_{\rm{diss}}\rm{(X_{2}O)}$ $\approx$~5.4~eV \cite{DK3}), leading to the following equation:

\begin{equation}\label{eq3}
\frac{1}{2} m_{\rm{OX}} v^{2}_{\rm{OX}}=\frac{E}{(1 + \frac{m_{\rm{OX}}}{m_{\rm{X}}})}
\end{equation}\\

Thus,  if the molecule is isolated and dissociates, the OD radicals  will be  formed  with higher initial translational energy than the OH radicals   (E$_{\rm{OD}}$=$E$/10 and    E$_{\rm{OH}}$=$E$/18, which leads to   $E_{\rm{OD}}$ $\approx$ 1.8$ \times E_{\rm{OH}}$) according to  Eq.~\ref{eq3}.
 Because the OD radicals have a higher initial translational energy than the OH radicals in the uppermost monolayers, they leave the ice surface more easily.

The OD photodesorption probability decreases with increasing depth, similar to the OH photodesorption probability \cite{Andersson2008, Arasa2010}. 
In the third and fourth MLs, the OD and OH photodesorption probabilities drop to less than 1$\times$10$^{-2}$  
because the OD and OH radicals do not have enough translational energy to escape from the ice surface.

 An oscillatory effect is observed when the OD and OH photodesorption probabilities are plotted in MLs 1--4 versus ice temperature (Fig.~\ref{Figure4}). We attributed these oscillations in our previous paper \cite{Arasa2010} to the irregular nature of the amorphous ice surface, which makes it very complicated to assign molecules to specific MLs, and also to the finite sample size of about 30 molecules per ML.

The average OD and OH photodesorption probabilities increase with ice temperature by  $\sim$24~$\%$ and   $\sim$25~$\%$, respectively,  from 10 to 90~K. 
If longer time scales in our simulations could be considered, a stronger dependence on ice temperature would be expected, because   processes like thermal  diffusion and thermal desorption are more efficient at higher $T_{\rm{ice}}$ \cite{Arasa2010}.

\subsection{D$_{2}$O molecule photodesorption}
\label{ssec:D2O}

\subsubsection{Kick-out $vs.$ direct mechanism}

The third photodesorption channel  upon UV photodissociation of D$_{2}$O amorphous ice  is D$_{2}$O molecule photodesorption.
This is  due to two mechanisms: the direct  and the kick-out mechanism.
The direct mechanism consists of the recombination of the D atom and OD radical that leads to the formation of an energetic D$_{2}$O molecule that eventually desorbs (outcome 5). In this situation the D$_{2}$O molecule has a high probability of desorbing from the ice in a vibrationally excited state \cite{Andersson2008}.
 The kick-out mechanism takes place after the photodissociation of a D$_{2}$O molecule when an energetic D atom transfers momentum to one of the surrounding D$_{2}$O molecules,  which is then likely to desorb vibrationally cold \cite{Yabushita1, Hama2010, Arasa2010}.
Since in our model we cannot quantify the energy transfer from X to internal modes of  X$_{2}$O,  because the kicked out molecule is treated as internally rigid,  
we have carried out  quasi-classical trajectory calculations on  the isolated X--X$_{2}$O   system at  different incident X atom kinetic energies (X=H, D).  Here we only report the results for  $E_{\rm{X}}$=1.5~eV,  because it is the average   kinetic energy with which H and D atoms kick a surrounding molecule out of  the ice at $T_{\rm{ice}}$=10~K.  Further details on these calculations will be reported in the  paper  we are preparing  \cite{inpreparation}. 
Gas phase  collisions at   $E_{\rm{X}}$=1.5~eV  lead to  final average vibrational energies of $\sim$0.16~eV for  H$_{2}$O, and $\sim$0.77~eV for  D$_{2}$O,  whereas in the ice system recombined H$_{2}$O and D$_{2}$O molecules desorb from the ice surface with average   ro-vibrational energies of 5.3 and 5.4~eV, respectively.
Thus, the kicked out molecules are  much  more likely to be formed  in states with lower  vibrational  energies than the molecules formed after recombination of the photofragments.  
Yabushita $et~al.$  \cite{Yabushita1}, and Hama $et~al.$  \cite{Hama2010}  observed photodesorbed X$_{2}$O  in the ground vibrational state by using resonance-enhanced multiphoton ionization (REMPI) detection methods  and classified them as kicked out.  
They speculated  that when the X atom  kicks out a X$_{2}$O molecule, most of the energy is transferred into translation and much less into internal energy  because  the X atom hits the X$_{2}$O  molecule close to the center of mass.  This is supported  by our simulations (e.g., $\sim$89~$\%$ of the trajectories classified as kicked out occurs when the X atom kicks the X$_{2}$O molecule close to the oxygen atom at   $T_{\rm{ice}}$=10~K)  (X=H, D).
Isotope effects are also observed in our calculations:  the efficiency of energy transfer from  D  to intramolecular vibrational modes of D$_{2}$O molecule in gas phase collisions is larger than the corresponding efficiency for collisions of  H with  H$_{2}$O, because D is heavier than H  and because the vibrational frequencies of   D$_{2}$O are lower than those of H$_{2}$O.

Fig.~\ref{Figure5}(a) shows the average  of the  D$_{2}$O photodesorption probabilities  compared with those for   H$_{2}$O  \cite{Arasa2010} over the top four monolayers versus ice temperature.  The average values for D$_{2}$O and H$_{2}$O \cite{Arasa2010}  due to the  kick-out mechanism are displayed in Fig.~\ref{Figure5}(b), and  those due to the direct mechanism  in Fig.~\ref{Figure5}(c).

The total average D$_{2}$O photodesorption probabilities are larger than those for H$_{2}$O  (Fig.~\ref{Figure5}(a)) because of the large contribution of the D$_{2}$O kick-out mechanism  (Fig.~\ref{Figure5}(b)).
For every ice temperature studied, the D$_{2}$O  photodesorption probability due to the kick-out mechanism is larger than that for  H$_{2}$O  (Fig.~\ref{Figure5}(b)), because in the case of heavy water, the D atom formed after D$_{2}$O photodissociation has  a larger momentum (by about a factor  $\sqrt{2}$)  than the  H atom formed after H$_{2}$O photodissociation. In addition,  the efficiency of momentum transfer from D to D$_{2}$O is larger than that from H to H$_{2}$O,  so that the kick-out mechanism is much  more successful for D$_{2}$O. The effect occurs mainly after photoexcitation in the second and third monolayers (see Figs.~\ref{Figure6}(b) and \ref{Figure6}(c)), because a D atom produced in these MLs  is more likely to kick-out a molecule located above it \cite{Andersson2006, Andersson2008,Arasa2010}.
The kick-out photodesorption probabilities are much lower in the first and fourth MLs  (Figs.~\ref{Figure6}(a) and \ref{Figure6}(d)), where  there are no important differences between  D$_{2}$O and H$_{2}$O.

We have  calculated the probabilities of the parallel outcomes that take place in coincidence with the kicking out of a D$_{2}$O molecule by an energetic D atom. These probabilities are summarized in Table~\ref{Table1}. We do not observe differences with those  calculated for the H$_{2}$O kicked out molecules from H$_{2}$O ice (Table~I in Ref.~\cite{Arasa2010}). The most dominant simultaneous process is that where the D atom that kicks out the D$_{2}$O molecule also desorbs, while the OD fragment is trapped (Table~\ref{Table1}). 
The next most important parallel processes are those in which the  photofragments recombine and form a D$_{2}$O molecule that  remains trapped in the ice, and those in which both photofragments are trapped inside the ice at separate locations. 
It is also possible that two molecules  desorb at the same time, i.e.,  the kicked out molecule and a recombined molecule, but this process occurs with  very low probability.

\begin{table}[htbp!]
\caption{Probabilities averaged over the top four  monolayers of the outcomes that take place in coincidence with the kicking out of a D$_{2}$O molecule for each ice temperature.
Overall probabilities can be obtained by multiplying the probabilities shown with the probabilities for the kick-out mechanism, see Fig.~\ref{Figure5}(b).}
\begin{center}
\begin{tabular} { c || c | c | c | c | c| c} 
\hline
\hline
$T_{\rm{ice}}$ / K  & D$_{\rm{des}}$ + OD$_{\rm{trap}}$ & D$_{\rm{des}}$ + OD$_{\rm{des}}$  & D$_{2}$O$_{\rm{des}}$ & D$_{\rm{trap}}$ + OD$_{\rm{trap}}$ & D$_{2}$O$_{\rm{trap}}$ & Others \\ 

  &  &  $\times$10$^{-3}$ & $\times$10$^{-3}$ &  &  &  $\times$10$^{-3}$\\ \hline

10 & 0.487 $\pm$ 0.047 & 0 & 0 & 0.257 $\pm$ 0.041 & 0.257 $\pm$ 0.041 & 0 \\
20 & 0.412 $\pm$ 0.040 & 6.8 $\pm$ 6.7 & 6.8 $\pm$ 6.7 & 0.108 $\pm$ 0.026 & 0.459 $\pm$ 0.041 &  6.8 $\pm$ 6.7 \\
60 & 0.511 $\pm$ 0.031 & 0 & 0 & 0.211 $\pm$ 0.025 & 0.278 $\pm$ 0.027 &  0 \\
90   & 0.587 $\pm$ 0.028 & 3.2 $\pm$ 3.2   & 3.2 $\pm$ 3.2 & 0.167 $\pm$ 0.021 & 0.237 $\pm$ 0.024 & 3.2 $\pm$ 3.2 \\ \hline
\end{tabular}

\end{center} 
\label{Table1}

\end{table}

\subsubsection{Trends with ice temperature}

\begin{figure}[htbp]
\begin{center}
\includegraphics[width=16cm]{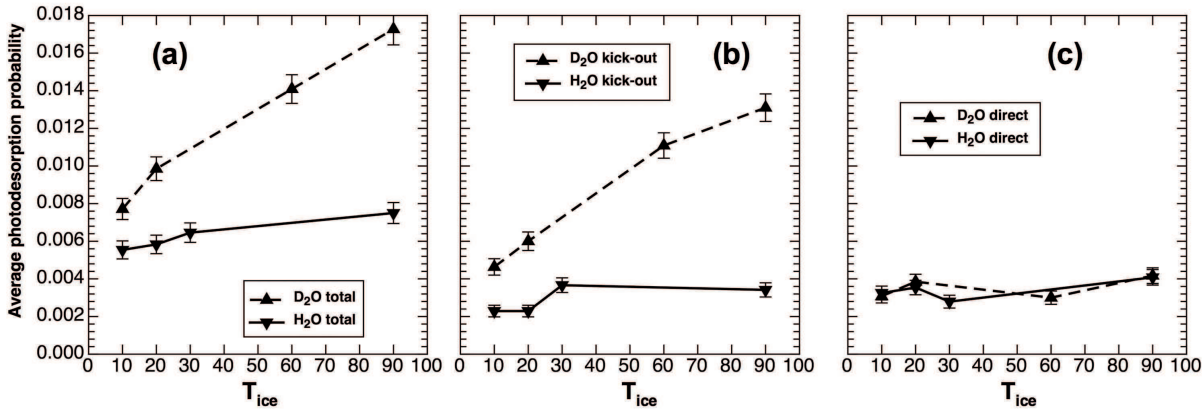}
\end{center}
\caption {{(a) The total D$_{2}$O and H$_{2}$O \cite{Arasa2010} photodesorption probabilities, (b) 
the D$_{2}$O and H$_{2}$O  \cite{Arasa2010} photodesorption probabilities due to the kick-out mechanism, and  (c) the probabilities due to the direct mechanism (per absorbed UV photon) versus ice temperature, all averaged over the top four MLs in which the photoexcited molecule resides.}}
\label{Figure5}
\end{figure}

The total D$_{2}$O photodesorption probability increases faster with ice temperature than that for H$_{2}$O (Fig.~\ref{Figure5}(a)): by 130~$\%$ $vs$ 30~$\%$ \cite{Arasa2010}, going from 10 to 90~K.
The average desorption probability due to the direct mechanism (Fig.~\ref{Figure5}(c)) is relatively small, and
there are no differences between the average D$_{2}$O and H$_{2}$O photodesorption probabilities: both  rise with ice temperature by only $\sim$30~$\%$ going from 10 to 90~K. 
However,  the D$_{2}$O average kick-out  photodesorption probability increases strongly with ice temperature (by $\sim$180~$\%$ from 10 to 90~K), whereas the H$_{2}$O kick-out average photodesorption probability shows a much  weaker increase with ice temperature (by $\sim$50~$\%$ from 10 to 90~K).  
Thus, the stronger  trend with ice temperature for   D$_{2}$O results from  the  increase of the D$_{2}$O kick-out photodesorption probability (Fig.~\ref{Figure5}(b)). 
This probability increases with ice temperature  because the molecules have higher initial kinetic energies at higher ice temperatures, which promotes the desorption of the surrounding molecules through the kick-out mechanism.

\begin{figure}[htbp]
\begin{center}
\includegraphics[width=18cm]{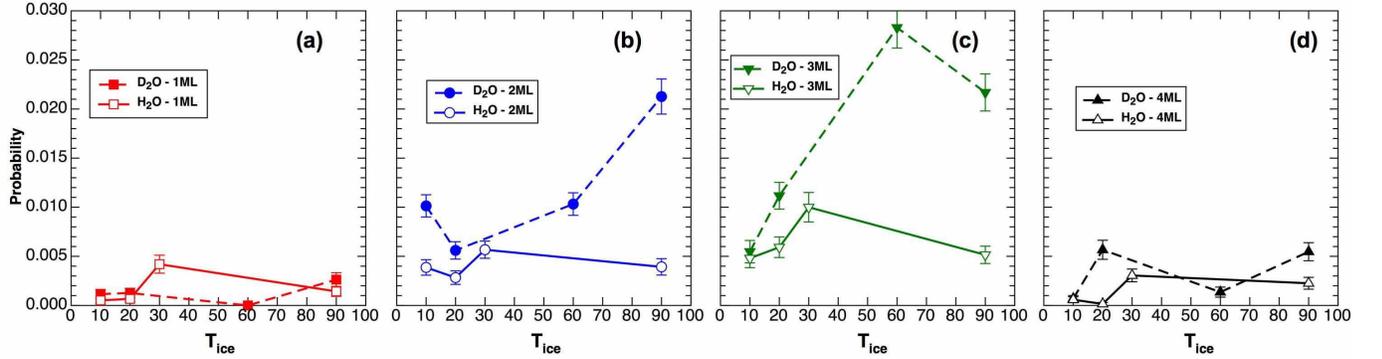}
\end{center}
\caption {{  Probabilities of D$_{2}$O molecule (dashed line) and H$_{2}$O molecule (solid line) \cite{Arasa2010} photodesorption due to the kick-out mechanism upon photoexcitation in (a) the first ML, (b) the second ML, (c) the third ML, and (d) the fourth ML, (per absorbed UV photon) versus ice temperature.}}
\label{Figure6}
\end{figure}

At higher ice temperatures the D$_{2}$O kick-out mechanism has a high probability (per absorbed UV photon):~2.8~$\%$ at 60~K after photoexcitation in the 3rd~ML (Fig.~\ref{Figure6}(c)) and 2.2~$\%$ at 90~K after photoexcitation in the 2nd~ML (Fig.~\ref{Figure6}(b)).
The probabilities for D$_{2}$O and H$_{2}$O desorption through the kick-out mechanism in the top 4~MLs (Fig.~\ref{Figure6}) show an oscillatory dependence on ice temperature, as also  seen before for OD and OH photodesorption probabilities versus ice temperature and monolayer (Fig.~\ref{Figure4}). These oscillations are attributed to the corrugation of the amorphous ice surface (for more details see Ref.~\cite{Arasa2010}).

\subsection{Energies of the kicked out molecules}
\label{ssec:kicked}

\begin{figure}[htbp]
\begin{center}
\includegraphics[width=10cm]{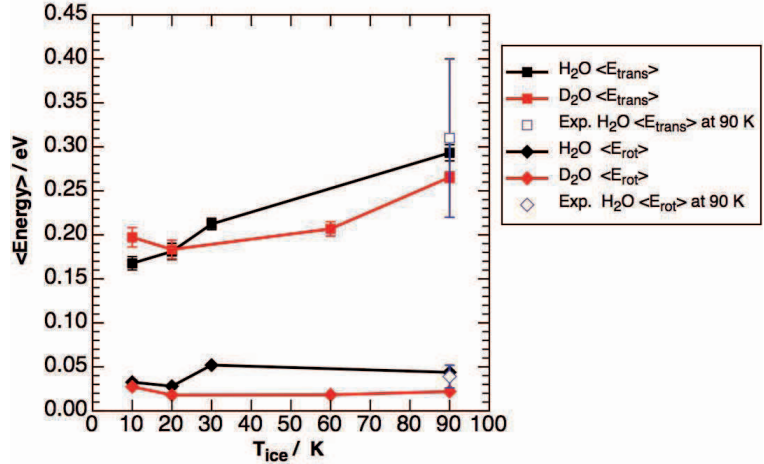}
\end{center}
\caption {{ Calculated average translational and  rotational energies  of the kicked out D$_{2}$O molecules versus ice temperature, of the kicked out H$_{2}$O molecules \cite{Arasa2010}, and experimental average translational and rotational energies of  H$_{2}$O molecules desorbed in their ground vibrational state at $T_{\rm{ice}}$=90~K \cite{Yabushita1}. }}
\label{Figure8}
\end{figure}

The average translational and rotational energies taken over the top 4 MLs of the  D$_{2}$O and H$_{2}$O \cite{Arasa2010} kicked out molecules are plotted versus ice temperature in  Fig.~\ref{Figure8}, together with the corresponding experimental values for H$_{2}$O photodesorbed molecules  ($v$=0) at 90~K \cite{Yabushita1}.
The translational energies tend to increase with ice temperature, because the energy of the ice also rises. 
The final rotational energies are low and do not display any dependence on ice temperature. The calculated translational and rotational energies do not show a significant isotope effect. 
Our calculations cannot say anything about the vibrational state of the kicked out molecules, because the molecules that are not photoexcited are kept rigid in our model. However, it seems unlikely that the kicked out molecules would emerge highly  vibrationally excited.
The average translational and rotational energies at 90~K of the  H$_{2}$O kicked out molecules are 0.29 and 0.044~eV, respectively \cite{Arasa2010}, and of the  D$_{2}$O kicked out molecules  0.27 and 0.021~eV, respectively. 
These results are in good agreement with the experimental translational and rotational  energies of the H$_{2}$O desorbed  molecules in their ground vibrational state as measured by Yabushita $et~al.$ \cite{Yabushita1} at 90~K  (0.31 and 0.039~eV, respectively), and also with the experimental translational and rotational
energies measured by Hama $et~al.$ \cite{Hama2010} for H$_{2}$O and D$_{2}$O ices at 90~K  (0.31 and 0.047~eV, respectively). Hama $et~al.$ \cite{Hama2010} did not observe differences between the energies of desorbed  H$_{2}$O ($v$=0) and desorbed  D$_{2}$O ($v$=0) at 90~K, in agreement with our calculations.

\subsection{Total (OD + D$_{2}$O) photodesorption yield and comparison with experiments}
\label{ssec:Total}

\begin{figure}[htbp!]
\begin{center}
\includegraphics[width=8cm]{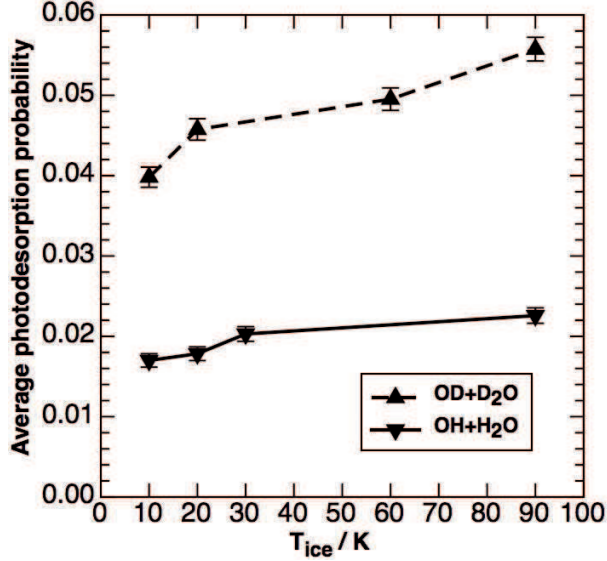}
\end{center}
\caption {{ The probability of (OD$_{\rm{des}}$ + D$_{2}$O$_{\rm{des}}$)  photodesorption (dashed line) and  (OH$_{\rm{des}}$ + H$_{2}$O$_{\rm{des}}$) photodesorption (solid line)  averaged over the top four MLs per absorbed UV photon is shown as a function of ice temperature. The  (OH$_{\rm{des}}$ + H$_{2}$O$_{\rm{des}}$) photodesorption results are taken from Ref.~\cite{Arasa2010}.}}
\label{Figure9}
\end{figure}

The calculated average of the total (OX + X$_{2}$O, for X=H or D) photodesorption probability per absorbed UV photon  is larger for D$_{2}$O ice than for H$_{2}$O ice at every ice temperature (Fig.~\ref{Figure9}).

In our previous study \cite{Arasa2010}, we compared the total photodesorption yield of the O containing species (H$_{2}$O$_{\rm{des}}$ + OH$_{\rm{des}}$) with the total experimental photodesorption yield (Eq.~4 in {\"O}berg $et~al.$ \cite{Oberg1}). In the calculation of this  yield, the deuterium (or hydrogen) atom photodesorption is not included (outcomes 2,  3, 5, and the kick-out), because D (or H) was not detected in the experiments by {\"O}berg $et~al.$  \cite{Oberg1}. 
The photodesorption yields   (i.e., photodesorption probabilities per incident photon) have been calculated through  Eq.~\ref{eq2} (see Sec.~\ref{ssec:Initial}).

Although the actual experimental photodesorption yields (Table~\ref{Table2}) \cite{Oberg1} were for D$_{2}$O ice, {\"O}berg $et~al.$ \cite{Oberg1} applied the results for  D$_{2}$O ice to H$_{2}$O ice (Eq.~4 in Ref.~\cite{Oberg1}), because they found that the total  photodesorption yields from D$_{2}$O and H$_{2}$O were indistinguishable (i.e., no isotope effects) within the experimental uncertainties (60$\%$) at 18 and 100~K.

Table~\ref{Table2}  contains the total experimental photodesorption yield \cite{Oberg1}, the computed total (H$_{2}$O$_{\rm{des}}$ + OH$_{\rm{des}}$)  \cite{Arasa2010}, and the computed total (D$_{2}$O$_{\rm{des}}$ + OD$_{\rm{des}}$) photodesorption yield  per incident photon,  the ratio $\xi$ between the experimental yield and the calculated (H$_{2}$O$_{\rm{des}}$ + OH$_{\rm{des}}$) yield, and  the same for the (D$_{2}$O$_{\rm{des}}$ + OD$_{\rm{des}}$)  yield,  at all ice temperatures.  
$\xi$ increases from 3.0 to 5.9 for  H$_{2}$O ice and from 1.3 to 2.3 for D$_{2}$O ice. 
Thus, the computed (D$_{2}$O$_{\rm{des}}$ + OD$_{\rm{des}}$) photodesorption yield is significantly larger than the computed (H$_{2}$O$_{\rm{des}}$ + OH$_{\rm{des}}$) yield at all ice temperatures, as also illustrated  in  Fig.~\ref{Figure9}, and compares better with the experimental D$_{2}$O  photodesorption yield than the yield  calculated for amorphous H$_{2}$O ice \cite{Arasa2010}.

The  agreement between theory and experiment is better at low ice temperatures,  (e.g., $\xi$=1.3 at 10~K and $\xi$=2.3 at 90~K). 
 This trend strengthens our explanation that the difference between theory and experiments can be due  to long time scale processes promoted by prolonged irradiation effects leading to an accumulation of radicals,  thermal desorption and thermal diffusion. 
At higher ice temperatures, the photofragments, possibly formed by different photodissociation events, become more  mobile, allowing them to recombine and eventually desorb as a consequence of the excess  energy.
Some of the OD photofragments that are trapped deeper in the ice, could probably desorb at long time scales due to a higher diffusion rate at higher $T_{\rm{ice}}$.
In our simulations we can only reach the picosecond time scale, therefore these kind of secondary processes  are beyond the scope of our simulations \cite{Arasa2010}.
 Another difference is  the UV wavelength covered by the lamp  used in the experiments \cite{Oberg1}. This UV   lamp includes Lyman-$\alpha$ photons which can excite  H$_{2}$O to the $\rm{\tilde{B}}$  state  whereas our calculations consider only the $\rm{\tilde{A}}$ state \cite{Andersson2006}.
Given the experimental uncertainties and our approximations (such as the use of a gas phase PESs for the H$_{2}$O intramolecular interactions, the freezing of the intramolecular degrees of freedom of the surrounding molecules, and  the short time scale of our simulations \cite{Andersson2006, Arasa2010}), the experimental and calculated probabilities may be considered to be in  reasonable  agreement. 
An important result for astrochemists is that the computational results fall within the range of the photodesorption probability per incident photon (1$\times$10$^{-4}$--3.5$\times$10$^{-3}$) \cite{Bergin1995, Willacy200, Snell2005, Dominik2005, Bergin2005} used to model astrophysical environments. 
However, our calculations suggest that the computed total photodesorption yield can be different for H$_{2}$O and D$_{2}$O ice, in contrast to the experimental results.

\begin{table}[htbp!]
\caption{Experimental $^\mathrm{a}$, theoretical (OH$_{\rm{des}}$ + H$_{2}$O$_{\rm{des}}$) \cite{Arasa2010}, and theoretical (OD$_{\rm{des}}$ + D$_{2}$O$_{\rm{des}}$) photodesorption yields per incident photon, the experimental yield/theoretical yield $\xi$ (H$_{2}$O) and the experimental yield/theoretical yield $\xi$ (D$_{2}$O) at all ice temperatures.}
\begin{center}
\begin{tabular} { c   || c | c | c | c | c } 
\hline
\hline
$T_{\rm{ice}}$ / K & Exp. (OX$_{\rm{des}}$ + X$_{2}$O$_{\rm{des}}$), X=H or D   & (OH$_{\rm{des}}$ + H$_{2}$O$_{\rm{des}}$) & (OD$_{\rm{des}}$ + D$_{2}$O$_{\rm{des}}$) & $\xi$ (H$_{2}$O)  & $\xi$ (D$_{2}$O) \\ 

 & $\times$10$^{-3}$   & $\times$10$^{-3}$ & $\times$10$^{-3}$ &   &  \\ \hline

10   & 1.62 $\pm$ 0.48 & 0.54 $\pm$ 0.04 & 1.27 $\pm$ 0.063 & 3.0 &  1.3 \\
20   & 1.94 $\pm$ 0.56 & 0.57 $\pm$ 0.04 & 1.46 $\pm$ 0.066 & 3.4 &  1.3 \\  
30   & 2.26 $\pm$ 0.64 & 0.71 $\pm$ 0.09 &    & 3.2 &   \\
60   & 3.22 $\pm$ 0.88 &  & 1.57 $\pm$ 0.068   &  &  2.1 \\
90   & 4.18 $\pm$ 1.1 & 0.71 $\pm$ 0.05 & 1.83 $\pm$ 0.074 &   5.9 & 2.3  \\ \hline
\end{tabular}

\begin{list} {} {}
\item $^\mathrm{a}$ Calculated from the empirical fit of the total photodesorption yield, Eq.~4 in {\"O}berg $et~al.$ \cite{Oberg1}.
\end{list}

\end{center} 

\label{Table2}
\end{table}

\section{Conclusions}
\label{sec:conclusions}

In this work we have studied the processes following UV photodissociation of D$_{2}$O in  amorphous heavy water ice and compared them with previous UV photodissociation results in H$_{2}$O ice at different ice temperatures,  in order to investigate  isotope effects in photodesorption.

D  atom photodesorption is the most important desorption process in the uppermost MLs of the ice, like H atom photodesorption. 
The average D atom photodesorption probability is smaller than that of the H atom,  because in the top two MLs of the ice both the H and D atom can easily escape from the ice surface, but if the atoms are located in the third and fourth MLs,
the D atom is less likely to penetrate the upper ice layers due to more efficient collision energy transfer to D$_{2}$O. Therefore,   the D atom photodesorption probabilities in these MLs  decrease, and the same is then for the average D atom photodesorption probability.
The D and H atom photodesorption probabilities do not show any dependence on ice temperature.

OD and OH radical photodesorption constitute the second most important desorption channel in the top two MLs of the ice. Deeper into the ice the probabilities of these processes drop because OD and OH radicals do not have enough translational energy to desorb from the surface.  The average OD  photodesorption probability is higher than that of OH.  
This trend can be explained by the initial translational energy of OD being higher by about a factor 1.8, 
 a result obtained if the photoexcited molecule is considered to be isolated (i.e., in the absence of the surrounding ice) and the laws of momentum and energy conservation are applied. 
The average OD photodesorption probabilities increase smoothly with ice temperature, by about 24~$\%$ from 10 to 90~K.

The third most important desorption mechanism  is D$_{2}$O and H$_{2}$O photodesorption. This process takes place either by direct desorption of the photoexcited molecule after the recombination of D (H) and OD (OH)  or by indirect desorption due to an energetic D (H) atom which transfers part of its momentum to a surrounding molecule that is kicked out from the ice surface.  
The  average  photodesorption probability  is higher for D$_{2}$O than  for H$_{2}$O  at all ice temperatures considered.  This trend is due to the contribution of the kick-out mechanism, 
which  is much more important for D$_{2}$O than for  H$_{2}$O.
 This result is expected because the photoproduced  D atoms have higher average momentum (by about a factor $\sqrt{2}$) than the H atoms, and because energy transfer in D--D$_{2}$O collisions is more efficient than energy transfer in H--H$_{2}$O collisions.
 The kick-out mechanism mainly takes place when the photoexcited molecule is initially located in the second and third MLs of the ice.
Photodissociation   leads to an energetic D (H) atom that can transfer its momentum to a molecule located above it, which will desorb from the ice if it has enough kinetic energy.
The average direct photodesorption probability (which involves recombination) does not show any  isotope effect. The average total D$_{2}$O photodesorption probability tends to increase with ice temperature faster than that of H$_{2}$O: by $\sim$130~$\%$ $vs$ $\sim$30~$\%$,  from 10 to 90~K.

Experiments show, and a consideration of the mechanism suggests,  that the 
 kicked out molecules leave the surface vibrationally cold.  In contrast, the molecules that desorb due to the direct mechanism are formed vibrationally excited.	The average translational and rotational energies in which the D$_{2}$O and H$_{2}$O molecules desorb due to the kick-out mechanism have been calculated and compared with the corresponding  experimental values at 90~K. The agreement between our MD calculations and the experimental measurements is good, and leads to the conclusion that the final energies with which the kicked out molecules are formed do not display an isotope effect.

We have also estimated the total photodesorption probability (OD$_{\rm{des}}$ + D$_{2}$O$_{\rm{des}}$) per incident photon from the total photodesorption probabilities per absorbed UV photon,  and compared this quantity with the  previously  calculated values for (OH$_{\rm{des}}$ + H$_{2}$O$_{\rm{des}}$),  and with the available experimental yields. 
Our total  photodesorption probability for D$_{2}$O compares  better with the experimental photodesorption yield than that for H$_{2}$O, and also better at low ice temperatures.
Presumably  at higher ice temperatures long time scale processes become increasingly important, such as diffusion and thermal desorption, which are not covered in our picosecond simulations.

Current experiments cannot distinguish between (OH + H$_{2}$O) and  (OD + D$_{2}$O) yields within the experimental uncertainties of 60$\%$. More accurate future  experiments may reveal the isotope effects predicted here.

 \begin{acknowledgments}

 The authors would like to thank M. C. van Hemert and   T. P. M. Goumans for valuable discussions.  This project was funded with  computer time by NCF/NWO, and with  TOP grant No. 700.56.321 by CW/NWO.
 
\end{acknowledgments}


\begin{thebibliography}{10}

\bibitem{Herbst2009}
E. Herbst and E.~F. van Dishoeck, Annu. Rev. Astron. Astrophys. {\bf \bf 47},
  427  (2009).

\bibitem{Tielens1991}
A.~G. G.~M. Tielens, A.~T. Tokunaga, T.~R. Geballe, and F. Baas, Astrophys. J.
  {\bf \bf 381},  181  (1991).

\bibitem{Boogert2008}
A.~C.~A. Boogert, K.~M. Pontoppidan, C. Knez, F. Lahuis, J. Kessler-Silacci,
  E.~F. van Dishoeck, G.~A. Blake, J.-C. Augereau, S.~E. Bisschop, S.
  Bottinelli, T.~Y. Brooke, J. Brown, A. Crapsi, N.~J. Evans~II, H.~J. Fraser,
  V. Geers, T.~L. Huard, J.~K. J$\o$rgensen, K.~I. {\"O}berg, L.~E. Allen,
  P.~M. Harvey, D.~W. Koerner, L.~G. Mundy, D.~L. Padgett, A.~I. Sargent, and
  K.~R. Stapelfeldt, Astrophys. J. {\bf \bf 678},  985  (2008).

\bibitem{Tanaka1994}
M. Tanaka, T. Nagata, S. Sato, and T. Yamamoto, Astrophys. J. {\bf \bf 430},
  779  (1994).

\bibitem{Chiar1995}
J.~E. Chiar, A.~J. Adamson, T.~H. Kerr, and D.~C.~B. Whittet, Astrophys. J.
  {\bf \bf 455},  234  (1995).

\bibitem{Gibb2000}
E.~L. Gibb, D.~C.~B. Whittet, W.~A. Schutte, A.~C.~A. Boogert, J.~E. Chiar, P.
  Ehrenfreund, P.~A. Gerakines, J.~V. Keane, A.~G. G.~M. Tielens, E.~F. van
  Dishoeck, and O. Kerkhof, Astrophys. J. {\bf \bf 536},  347  (2000).

\bibitem{Gibb2004}
E.~L. Gibb, D.~C.~B. Whittet, A.~C.~A. Boogert, and A.~G. G.~M. Tielens,
  Astrophys. J. Sup. Series {\bf \bf 151},  35  (2004).

\bibitem{Pontoppidan2003}
K.~M. Pontoppidan, H.~J. Fraser, E. Dartois, W.-F. Thi, E.~F. van Dishoeck,
  A.~C.~A. Boogert, L. d'Hendecourt, A.~G. G.~M. Tielens, and S.~E. Bisschop,
  Astron. Astrophys. {\bf \bf 408},  981  (2003).

\bibitem{Pontoppidan2006}
K.~M. Pontoppidan, Astron. Astrophys. {\bf \bf 453},  L47  (2006).

\bibitem{Zasowski2009}
G. Zasowski, F. Kemper, D.~M. Watson, E. Furlan, C.~J. Bohac, C. Hull, and
  J.~D. Green, Astrophys. J. {\bf \bf 694},  459  (2009).

\bibitem{Butner2007}
H.~M. Butner, S.~B. Charnley, C. Ceccarelli, S.~D. Rodgers, J.~R. Pardo, B.
  Parise, J. Cernicharo, and G.~R. Davis, Astrophys. J. {\bf \bf 659},  L137
  (2007).

\bibitem{Vastel2010}
C. {Vastel}, C. {Ceccarelli}, E. {Caux}, A. {Coutens}, J. {Cernicharo}, S.
  {Bottinelli}, K. {Demyk}, A. {Faure}, L. {Wiesenfeld}, Y. {Scribano}, A.
  {Bacmann}, P. {Hily-Blant}, S. {Maret}, A. {Walters}, E.~A. {Bergin}, G.~A.
  {Blake}, A. {Castets}, N. {Crimier}, C. {Dominik}, P. {Encrenaz}, M.
  {G{\'e}rin}, P. {Hennebelle}, C. {Kahane}, A. {Klotz}, G. {Melnick}, L.
  {Pagani}, B. {Parise}, P. {Schilke}, V. {Wakelam}, A. {Baudry}, T. {Bell}, M.
  {Benedettini}, A. {Boogert}, S. {Cabrit}, P. {Caselli}, C. {Codella}, C.
  {Comito}, E. {Falgarone}, A. {Fuente}, P.~F. {Goldsmith}, F. {Helmich}, T.
  {Henning}, E. {Herbst}, T. {Jacq}, M. {Kama}, W. {Langer}, B. {Lefloch}, D.
  {Lis}, S. {Lord}, A. {Lorenzani}, D. {Neufeld}, B. {Nisini}, S. {Pacheco}, J.
  {Pearson}, T. {Phillips}, M. {Salez}, P. {Saraceno}, K. {Schuster}, X.
  {Tielens}, F. {van der Tak}, M.~H.~D. {van der Wiel}, S. {Viti}, F.
  {Wyrowski}, H. {Yorke}, P. {Cais}, J.~M. {Krieg}, M. {Olberg}, and L.
  {Ravera}, Astron. Astrophys. {\bf \bf 521},  L31  (2010).

\bibitem{Ehrenfreund2003}
P. Ehrenfreund and H. Fraser, {\em { Solid State Astrochemistry, {\em V.
  Pirronello, J. Krelowski (Eds.)}}} (Kluwer Academic Publishers, Dordrecht,
  2003), p.\ 317.

\bibitem{Garrod2006}
R.~T. Garrod and E. Herbst, Astron. Astrophys. {\bf \bf 457},  927  (2006).

\bibitem{Oberg1}
K.~I. {\"O}berg, H. Linnartz, R. Visser, and E.~F. van Dishoeck, Astrophys. J.
  {\bf \bf 693},  1209  (2009).

\bibitem{Boogert2004}
A.~C.~A. Boogert and P. Ehrenfreund, Astrophys. of Dust, ASP Conference Series
  {\bf \bf 309},  547  (2004).

\bibitem{Kobayashi1983}
K. Kobayashi, J. Chem. Phys. {\bf \bf 87},  4317  (1983).

\bibitem{vanDishoeck1988}
E.~F. van Dishoeck, {\em {{\em in} Rate Coefficients in Astrochemistry, {\em
  edited by T. J. Millar and D. A. Williams} }} (Kluwer Academic Publishers,
  Dordrecht, 1988), p.\ 49.

\bibitem{Jack1988}
T. Jacq, C. Henkel, C.~M. Walmsley, P.~R. Jewell, and A. Baudry, Astron.
  Astrophys. {\bf \bf 199},  L5  (1988).

\bibitem{Knacke1991}
R.~F. Knacke and H.~P. Larson, Astrophys. J. {\bf \bf 367},  162  (1991).

\bibitem{Cernicharo1990}
J. Cernicharo, C. Thum, H. Hein, D. John, P. Garcia, and F. Mattioco, Astron.
  Astrophys. {\bf \bf 231},  L15  (1990).

\bibitem{Gensheimer1996}
P.~D. Gensheimer, R. Mauersberger, and T.~L. Wilson, Astron. Astrophys. {\bf
  \bf 314},  281  (1996).

\bibitem{vanderTak2006}
F.~F.~S. van~der Tak, C.~M. Walmsley, F. Herpin, and C. Ceccarelli, Astron.
  Astrophys. {\bf \bf 447},  1011  (2006).

\bibitem{Kaufman2008}
M.~J. Kaufman, D.~J. Hollenbach, E. Bergin, and G.~J. Melnick, EAS Publications
  Series {\bf \bf 31},  43  (2008).

\bibitem{Hollenbach2004}
D. Hollenbach and U. Gorti, RevMexAA (Conference Series) {\bf \bf 22},  33
  (2004).


\bibitem{Hollenbach2009}
D. Hollenbach, M.~J. Kaufman, E.~A. Bergin, and G.~J. Melnick, Astrophys. J.
  {\bf \bf 690},  1497  (2009).

\bibitem{dHendecourt1982}
L.~B. d'Hendecourt, L.~J. Allamandola, F. Baas, and J.~M. Greenberg, Astron. Astrophys. {\bf
  \bf 109},  L12  (1982).

\bibitem{Westley1995a}
M.~S. Westley, R.~A. Baragiola, R.~E. Johnson, and G.~A. Baratta, Nature {\bf
  \bf 373},  405  (1995).

\bibitem{Yabushita1}
A. Yabushita, T. Hama, M. Yokoyama, M. Kawasaki, S. Andersson, R.~N. Dixon,
  M.~N.~R. Ashfold, and N. Watanabe, Astrophys. J. {\bf \bf 699},  L80  (2009).

\bibitem{Westley1995b}
M.~S. Westley, R.~A. Baragiola, R.~E. Johnson, and G.~A. Baratta, Planet. Space
  Sci. {\bf \bf 43},  1311  (1995).

\bibitem{Yabushita2006}
A. Yabushita, D. Kanda, N. Kawanaka, M. Kawasaki, and M.~N.~R. Ashfold, J.
  Chem. Phys. {\bf \bf 125},  3406  (2006).

\bibitem{Yabushita2008}
A. Yabushita, T. Hama, D. Iida, N. Kawanaka, M. Kawasaki, N. Watanabe, M.~N.~R.
  Ashfold, and H.-P. Loock, J. Chem. Phys. {\bf \bf 129},  044501  (2008).

\bibitem{HamaAug2009}
T. Hama, A. Yabushita, M. Yokoyama, M. Kawasaki, and S. Andersson, J. Chem.
  Phys. {\bf \bf 131},  054508  (2009).

\bibitem{HamaSet12009}
T. Hama, A. Yabushita, M. Yokoyama, M. Kawasaki, and N. Watanabe, J. Chem.
  Phys. {\bf \bf 131},  114511  (2009).

\bibitem{HamaSet22009}
T. Hama, A. Yabushita, M. Yokoyama, M. Kawasaki, and N. Watanabe, J. Chem.
  Phys. {\bf \bf 131},  114510  (2009).

\bibitem{Ghormley1971}
J.~A. Ghormley and C.~J. Hochanadel, J. Phys. Chem. {\bf \bf 75},  40  (1971).

\bibitem{Gerakines1996}
P.~A. Gerakines, W.~A. Schutte, and P. Ehrenfreund, Astron. Astrophys. {\bf \bf
  312},  289  (1996).

\bibitem{Hama2010}
T. Hama, M. Yokoyama, A. Yabushita, M. Kawasaki, S. Andersson, C.~M. Western,
  M.~N.~R. Ashfold, R.~N. Dixon, and N. Watanabe, J. Chem. Phys. {\bf \bf 132},
   164508  (2010).

\bibitem{Watanabe2000}
N. Watanabe, T. Horii, and A. Kouchi, Astrophys. J. {\bf \bf 541},  772
  (2000).

\bibitem{Andersson2005}
S. Andersson, G.-J. Kroes, and E.~F. van Dishoeck, Chem. Phys. Lett. {\bf \bf
  408},  415  (2005).

\bibitem{Andersson2006}
S. Andersson, A. Al-Halabi, G.-J. Kroes, and E.~F. van Dishoeck, J. Chem. Phys.
  {\bf \bf 124},  064715  (2006).

\bibitem{Andersson2008}
S. Andersson and E.~F. van Dishoeck, Astron. Astrophys. {\bf \bf 491},  907
  (2008).

\bibitem{Arasa2010}
C. Arasa, S. Andersson, H.~M. Cuppen, E.~F. van Dishoeck, and G.-J. Kroes, J.
  Chem. Phys. {\bf \bf 132},  184510  (2010).

\bibitem{TIP4P}
W.~L. Jorgensen, J. Chandrasekhar, J.~D. Madura, R.~W. Impey, and M.~L. Klein,
  J. Chem. Phys. {\bf \bf 79},  926  (1983).

\bibitem{Allen1987}
M.~P. Allen and D.~J. Tildesley, {\em { Computer Simulations of Liquids}}
  (Clarendon, Oxford, 1987).

\bibitem{Essmann1995}
U. Essmann and A. Geiger, J. Chem. Phys. {\bf \bf 103},  4678  (1995).

\bibitem{AlHalabi2004a}
A. Al-Halabi, E.~F. van Dishoeck, and G.-J. Kroes, J. Chem. Phys. {\bf \bf
  120},  3358  (2004).

\bibitem{AlHalabi2004b}
A. Al-Halabi, H.~F. Fraser, G.-J. Kroes, and E.~F. van Dishoeck, Astron.
  Astrophys. {\bf \bf 422},  777  (2004).

\bibitem{Wigner2}
R. van Harrevelt, M.~C. van Hemert, and G.~C. Schatz, J. Phys. Chem. A {\bf \bf
  105},  11480  (2001).

\bibitem{Wigner1}
R. Schinke, {\em { Photodissociation Dynamics}} (Cambridge University Press,
  Cambridge, 1993).

\bibitem{DK1}
A.~J. Dobbyn and P.~J. Knowles, unpublished.

\bibitem{DK2}
F.~J. Aoiz, L. Ba$\tilde{n}$ares, J.~F. Castillo, M. Brouard, W. Denzer, C.
  Vallance, P. Honvault, J.-M. Launay, A.~J. Dobbyn, and P.~J. Knowles, Phys. Rev.
  Lett. {\bf \bf 86},  1729  (2001).

\bibitem{DK3}
R. van Harrevelt and M.~C. van Hemert, J. Chem. Phys. {\bf \bf 114},  9453
  (2001).

\bibitem{inpreparation}
C. Arasa, M.~C. van Hemert, E.~F. van Dishoeck, and G.~J. Kroes, in preparation.

\bibitem{Bergin1995}
E.~A. Bergin, W.~D. Langer, and P.~F. Goldsmith, Astrophys. J. {\bf \bf 441},
  222  (1995).

\bibitem{Willacy200}
K. Willacy and W.~D. Langer, Astrophys. J. {\bf \bf 544},  903  (2000).

\bibitem{Snell2005}
R.~L. Snell, D. Hollenbach, J.~E. Howe, D.~A. Neufeld, M.~J. Kaufman, G.~J.
  Melnick, E.~A. Bergin, and Z. Wang, Astrophys. J. {\bf \bf 620},  758
  (2005).

\bibitem{Dominik2005}
C. Dominik, C. Ceccarelli, D. Hollenbach, and M. Kaufman, Astrophys. J. {\bf
  \bf 635},  L85  (2005).

\bibitem{Bergin2005}
E.~A. Bergin and G. Melnick, {\em { in Astrochemistry: recent successes and
  current challenges, {\em ed. D. C. Lis, G. A. Blake, and E. Herbst, IAU Symp.
  }}} (Cambridge University Press, Cambridge, 2005), Vol.~231, p.\ 309.

\end{thebibliography}
\end{document}